\documentclass[
  twocolumn,
  prl,
  showpacs,
  amsmath,
  amssymb,
  superscriptaddress,
  floatfix
]{revtex4}

\usepackage{bm}
\usepackage{graphicx}

\newcommand{\pa}{\partial}

\newcommand{\be}{\begin{equation}}
\newcommand{\e}{\end{equation}}
\newcommand{\beml}{\begin{subequations}}
\newcommand{\eml}{\end{subequations}}
\newcommand{\beq}{\begin{eqnarray}}
\newcommand{\eq}{\end{eqnarray}}
\newcommand{\ba}{\begin{array}}
\newcommand{\ea}{\end{array}}
\newcommand{\bpm}{\begin{pmatrix}}
\newcommand{\epm}{\end{pmatrix}}
\newcommand{\lt}{\left}
\newcommand{\rt}{\right}
\newcommand{\n}{\nonumber}
\newcommand{\la}{\langle}
\newcommand{\ra}{\rangle}

\newcommand{\bb}{\boldsymbol}

\newcommand{\Tr}{\mathop{\mathbf{Tr}}}

\DeclareMathOperator{\tr}{Tr}

\begin{document}

\title{Diffusion and criticality in undoped graphene with resonant scatterers}

\author{P.~M.~Ostrovsky}
\affiliation{
 Institut f\"ur Nanotechnologie, Karlsruhe Institute of Technology,
 76021 Karlsruhe, Germany
}
\affiliation{
 L.~D.~Landau Institute for Theoretical Physics RAS,
 119334 Moscow, Russia
}

\author{M.~Titov}
\affiliation{
 School of Engineering \& Physical Sciences, Heriot-Watt University,
 Edinburgh EH14 4AS, UK
}
\affiliation{
 Institut f\"ur Nanotechnologie, Karlsruhe Institute of Technology,
 76021 Karlsruhe, Germany
}

\author{S.~Bera}
\affiliation{
 Institut f\"ur Nanotechnologie, Karlsruhe Institute of Technology,
 76021 Karlsruhe, Germany
}
\affiliation{
 DFG Center for Functional Nanostructures,
 Karlsruhe Institute of Technology, 76128 Karlsruhe, Germany
}

\author{I.~V.~Gornyi}
\affiliation{
 Institut f\"ur Nanotechnologie, Karlsruhe Institute of Technology,
 76021 Karlsruhe, Germany
}
\affiliation{
 A.F.~Ioffe Physico-Technical Institute,
 194021 St.~Petersburg, Russia.
}
\affiliation{
 DFG Center for Functional Nanostructures,
 Karlsruhe Institute of Technology, 76128 Karlsruhe, Germany
}

\author{A.~D.~Mirlin}
\affiliation{
 Institut f\"ur Nanotechnologie, Karlsruhe Institute of Technology,
 76021 Karlsruhe, Germany
}
\affiliation{
 Inst. f\"ur Theorie der kondensierten Materie,
 Karlsruhe Institute of Technology, 76128 Karlsruhe, Germany
}
\affiliation{
 Petersburg Nuclear Physics Institute,
 188300 St.~Petersburg, Russia.
}
\affiliation{
 DFG Center for Functional Nanostructures,
 Karlsruhe Institute of Technology, 76128 Karlsruhe, Germany
}

\begin{abstract}
A general theory is developed to describe graphene
with arbitrary number of isolated impurities.   The theory provides a
basis for an efficient numerical analysis of the charge transport and
is applied to calculate the minimal conductivity, $\sigma$, of graphene
with resonant scatterers. In the case of smooth resonant impurities $\sigma$
grows logarithmically with increasing impurity concentration,
in agreement with  renormalization group analysis for the symmetry class DIII.
For vacancies (or strong on-site potential impurities)
$\sigma$ saturates  at a constant value that depends on the
vacancy distribution among two sublattices as expected  for the
symmetry class BDI.
\end{abstract}


\maketitle

Transport properties of graphene \cite{Novoselov04,Novoselov05,
  GuineaRMP} remain in the focus of intense
studies. It has been established, both theoretically
\cite{Katsnelson,Tworzydlo06Beenakker08rev}
and experimentally \cite{Morpurgo06,Miao07,Danneau08,Bolotin,Andrei}, that the
conductivity of short and wide
samples of ballistic graphene acquires
a minimal value of $4\times e^2/\pi h$ (the factor 4 reflecting the
spin and valley degeneracy) when the chemical potential is tuned
to a vicinity of the Dirac point.
The minimal conductivity of larger graphene flakes
is close to $4\times e^2/ h$ for the
majority of experimentally available samples  \cite{Novoselov05}.
The enhancement of the minimal conductivity is attributed to the
effect of disorder: graphene
near the Dirac point may conduct better when impurities are added
\cite{Titov07,Bardarson07,Schuessler09}.

Remarkably, the minimal conductivity of disordered graphene remains
essentially constant when the temperature $T$ is lowered by several orders
of magnitude, down to $30\:{\rm mK}$ \cite{Kim}. This is in contrast
with the behavior of conventional 2D systems with conductivity $\sigma \sim
e^2/h$: their $\sigma$ gets strongly suppressed with lowering $T$ due
to Anderson localization.
The absence of localization in graphene indicates that the dominant
disorder is either of  long-range character (and thus does not mix the
valleys) or preserves a chiral symmetry of the Hamiltonian
\cite{Ostrovsky06,Ostrovsky07b}. The
former possibility has been investigated in
Refs.~\cite{Ostrovsky07a,Ostrovsky07b,Bardarson07,Schuessler09}. In
this paper we explore the case of resonant impurities that preserve
the chiral symmetry ($C_z$ in terminology of
Ref.~\cite{Ostrovsky06}).

Resonant scatterers create "mid-gap" states directly at the
Dirac point, thus having a strong   impact on the minimal
conductivity. A natural example of a resonant scatterer is a strong
potential applied to a site of a graphene honeycomb lattice,
which is equivalent
to a vacancy. In this way, vacancies are effectively   created by
hydrogen ad-atoms or CH$_3$ molecules, which bind to a single carbon
atom in graphene and change its hybridization from $sp^2$ to $sp^3$
type. The resonant  character
of hydrogen ad-atoms was supported by the DFT analysis \cite{Lichtenstein}.
Resonant scatterers give linear (up to a logarithmic factor) dependence
of the graphene conductivity on electron density
\cite{Ostrovsky06,Stauber07}, consistent with
experimental observations. Recent experiments \cite{Elias09,Ni10}
confirmed  that a moderate concentration of
hydrogen adsorbates preserves all the salient features of transport in
graphene and provided evidence that resonant impurities determine the
graphene mobility.

The vacancies are not the only type of resonant impurities. For
instance, a smooth potential impurity,  which can be represented by
a scalar potential in the Dirac equation, is resonant provided  the
energy of the localized impurity state coincides with the Dirac
point. Recent works \cite{Bardarson09,Titov10} studied the effect of
resonant scalar impurities in the ballistic regime. It was shown, in
particular, that
each such impurity enhances the conductance of ballistic graphene  by a
value of the order of $e^2/h$.

In this Letter we develop a general
theory of transport in a system with arbitrary number $N$ of isolated
impurities.  The full counting statistics of the system is given by
a determinant of a matrix of size $N$.  The averaging over impurity
positions
can be performed numerically with a high efficiency.
We apply this theory to disordered graphene at the Dirac point,
focusing on two types of resonant scatterers:  scalar (smooth
potential) impurities and vacancies. This allows us to explore the
diffusive (or critical) regime that is established with increasing
concentration of impurities (or, equivalently, sample length).

We consider a graphene sample of the length $L$ and the width $W\gg L$,
which is described  by the Dirac Hamiltonian, $H = -i \hbar v
\bm{\sigma} \mathbf{\nabla} + V(\mathbf{r})$,  where $\bm{\sigma} =
(\sigma_x, \sigma_y)$ is the vector of Pauli matrices, $v$ is the
velocity, and $V(\mathbf{r})$ is an impurity potential that can mix
valleys and sublattices. Hereafter we set $\hbar v = 1$.

Metallic leads at $x < 0$ and $x > L$ are defined by adding a large chemical
potential, $\mu_\infty \to \infty$. Inside the sample, i.e. for $0 < x < L$, the
chemical potential is tuned to the Dirac point ($\mu = 0$).
The function $V(\mathbf{r})=\sum_{n=1}^N V_n(\mathbf{r})$ represents $N$
isolated scatterers.

To calculate the full counting statistics of electron transport we use the
matrix Green function approach \cite{Nazarov94, Ludwig, Titov10}. The Green
function in the retarded-advanced (RA) space satisfies the equation
\begin{equation}
 \label{keldysh}
 \begin{pmatrix}
   \mu - H + i0 & -\sigma_x \zeta \delta(x) \\
   -\sigma_x \zeta \delta(x-L) & \mu - H - i0
 \end{pmatrix} \mathcal{G}(\mathbf{r}, \mathbf{r}')
  = \delta(\mathbf{r}-\mathbf{r}'),
\end{equation}
where $\zeta = \sin(\phi/2)$ is the counting field and the chemical
potential $\mu$ equals 
$\mu_\infty$ in the leads and zero in the sample. It is important for
the subsequent analysis 
that the bare Green function $\mathcal{G}_0$, which solves Eq.~(\ref{keldysh})
at $V=0$, can be calculated analytically \cite{Titov10}.

Transport quantities (conductance, noise, and higher order cumulants)
are readily determined from the cumulant generating function $\mathcal{F}=\Tr
\ln \mathcal{G}^{-1}$, where the trace $\Tr$ includes the spatial coordinates as
well as RA, sublattice, and valley indices.
Below we are mostly concerned with the conductance determined by the
relation
\be
\label{conductance}
G=-(4e^2/h) \big(\pa^2 \mathcal{F}/\pa \phi^2 \big)_{\phi=0}.
\e
With the help of the Dyson equation, $\mathcal{G}^{-1} =\mathcal{G}_0^{-1} - V$,
we obtain
$\mathcal{F}=\mathcal{F}_0+\delta \mathcal{F}$, where $\delta \mathcal{F}=\Tr
\ln (1-V \mathcal{G}_0)$ describes the correction to the cumulant generating
function, $\mathcal{F}_0=\Tr \ln \mathcal{G}_0^{-1}=-W\phi^2/2\pi L$, of the
clean system \cite{Titov10}.

Our aim is to take advantage of the fact that impurities do not overlap and
to reduce the operator determinant in the definition of $\delta \mathcal{F}$
to the usual matrix determinant. Such a reduction is justified if the impurity
size is smaller
than both the distance between impurities and the Fermi wave length (the latter
is infinite at the Dirac point). The generating function can be rewritten in
the $N$-dimensional unfolded (impurity) space as
\be
\label{unfolding}
\delta \mathcal{F}=\Tr \ln (1-V \mathcal{G}_0 )= \Tr \ln (1-\hat{V}
\hat{\mathcal{G}}_0),
\e
where $\hat{V}=\textrm{diag}(V_1,V_2,\dots, V_N)$ and
all elements of the matrix $\hat{\mathcal{G}}_0$
are identical and equal to the operator $\mathcal{G}_0$.


We proceed with introducing the $T$-matrix operators $T_n = (1 - V_n g)^{-1}
V_n$, that account for multiple rescattering on individual impurities
\cite{Titov10}. Here $g$ stands for the Green function in an infinite graphene
plane. At the Dirac point we find
\be
g(\mathbf{r},\mathbf{r}') = -(i/2\pi)\,\bb{\sigma}\cdot
(\mathbf{r}-\mathbf{r}')/|\mathbf{r}-\mathbf{r}'|^2.
\e
Using the definition of the $T$-matrix, we rewrite $\delta \mathcal{F}$ as
\be
\label{calculation}
\delta \mathcal{F} = \Tr \ln \big[1-\hat{T} (\hat{\mathcal{G}}_0 - g) \big] +
\Tr \ln (1-\hat{V} g ),
\e
where $\hat{T}=\textrm{diag}(T_1,T_2,\dots, T_N)$ and $g$ is proportional
to the unit matrix in the unfolded space.

The last term in Eq.~(\ref{calculation}) does not depend on the source
field, $\phi$, 
and can be safely omitted. Taking the limit of point-like impurity we reduce
the operator product in the first term of Eq.~(\ref{calculation}) to
the standard matrix product 
with the result
\be
\label{main}
\delta \mathcal{F} =\tr \ln (1-\hat{T}\hat{\mathcal{G}}_\textrm{reg}),
\e
where $\hat{T}$ is the diagonal matrix of integrated impurity $T$-matrices
(describing $s$-wave scattering only)
and the elements of the matrix $\hat{\mathcal{G}}_\textrm{reg}$
in the unfolded space are given by
\be
\label{defGreg}
(\hat{\mathcal{G}}_\textrm{reg})_{nm}=
\begin{cases}
\mathcal{G}_0(\mathbf{r}_n,\mathbf{r}_m),& m\neq n, \\
\lim\limits_{\mathbf{r}\to \mathbf{r}_n}\!\!
\big[\mathcal{G}_0(\mathbf{r}_n,\mathbf{r})-g(\mathbf{r}_n,\mathbf{r})\big], &
m=n,
\end{cases}
\e
where $\mathbf{r}_n=(x_n,y_n)$ specify the positions of impurities.

Equation (\ref{main}) is one of the central results of the Letter. The
impurity-induced correction 
to the full-counting statistics is reduced to the finite-size matrix
determinant, which is completely 
defined by the impurity $T$-matrices in the $s$-wave channel and the bare
Green function.
The $T$-matrices can be found in a standard way from the solution of the
corresponding
single impurity scattering problem \cite{HentschelNovikovBasko}. The exact bare
Green
function in the rectangular geometry of Eq.~(\ref{keldysh}) has been
calculated in Ref.~\cite{Titov10}. 
It is given by the matrix product
\be
\label{resultG0}
\mathcal{G}_0(\mathbf{r}_n,\mathbf{r}_m)=\frac{i}{4L}U_{x_n}\Lambda \Sigma_z
e^{\Sigma_y(y_n-y_m)\phi/2L} R \Lambda U_{x_m}^{-1},
\e
where $\Sigma_{x,y,z}$ are the Pauli matrices in RA space and
\be
\n
U_x=\bpm \sin\frac{\phi(x-L)}{2L}& \cos \frac{\phi(x-L)}{2L}\\ i \cos\frac{\phi
x}{2L} & i\sin\frac{\phi x}{2L} \epm_{RA},
\quad
\Lambda=
\bpm 1& 0\\ 0 & \sigma_z \epm_{RA}.
\e
The matrix $R$ is acting in the sublattice space. In the limit
$W\gg L$ it simplifies to
\be
\label{ret}
R(\mathbf{r}_n,\mathbf{r}_m)=
\bpm \csc(z_n+z^*_m) & \csc(z_n-z_m)\\
\csc(z^*_n-z^*_m) & \csc(z^*_n+z_m)
\epm_{\sigma}, 
\e
where $z_n=\pi(x_n+i y_n)/2L$ and $\csc z = 1/\sin z$. The matrix $R$ coincides
with the retarded Green
function of the clean sample up to a factor $4iL$.

The expressions (\ref{resultG0}), (\ref{ret}) define the off-diagonal
elements of the matrix $\hat{\mathcal{G}}_{\textrm{reg}}$ 
in the unfolded space. The diagonal elements are found from
Eq.~(\ref{defGreg}) as 
\be
\label{diagonal}
(\hat{\mathcal{G}}_\textrm{reg})_{nn}=\frac{i}{4L} U_{x_n} \Lambda \Sigma_z
R_\mathrm{reg} \Lambda U_{x_n}^{-1},
\e
where $R_\mathrm{reg}(\mathbf{r})=\csc(\pi x/L) +\Sigma_y \sigma_y
\phi/\pi$. Diagonal part of the matrix $R_\mathrm{reg}$ is proportional to the
local density of states in a clean setup. Equation (\ref{diagonal})
completes the construction of the matrix $\hat{\mathcal{G}}_{\textrm{reg}}$.

The only input parameters for the general result (\ref{main}) are $T$-matrices
of individual impurities. They can be obtained by solving the single impurity
scattering problem. For smooth potential impurity, the $T$-matrix
mixes neither sublattices nor valleys and is given by the scattering
length, $T=\ell$. The length $\ell$ diverges if the
impurity potential fulfills the resonant condition \cite{Titov10}. In contrast,
the $T$-matrix for an on-site potential impurity projects on a one-dimensional
subspace, $T = \ell |u \ra \la u| = \ell(1\mp\tau_x\sigma_x
\pm\tau_y\sigma_y+\tau_z\sigma_z)/4$, where upper (lower) sign correspond to
the impurity in A (B) sublattice, respectively, and $\tau_\alpha$
are the Pauli matrices in the valley space \cite{HentschelNovikovBasko}. The
vacancy (i.e. infinitely strong on-site potential) corresponds to the limit
$\ell\to \infty$.

In general, the $T$ matrix of a resonant impurity is given by a divergent length
scale, $\ell$,
multiplied by a projection operator acting in the valley and sub-lattice space.
This enables
further simplification of the result (\ref{main}) by omitting the unity under
the logarithm
in the projected basis. Up to an arbitrary constant term, the resulting
generating function
can be cast in the form
\be
\label{resonant}
\delta \mathcal{F} =\tr \ln \hat{K} \hat{K}^\dagger,
\e
with a matrix $\hat K$ satisfying the identity $\hat K^\dagger(\phi) =
\hat K(-\phi)$. For resonant scalar impurities, the elements of $2N\times 2N$
matrix $\hat K$ are given by
\be
\label{scalar}
K_{nm}=
\begin{cases}
\sigma_z R(\mathbf{r}_n,\mathbf{r}_m), & m \neq n, \\
\sigma_z\csc(\pi x_n/L)-i\sigma_x \phi/\pi, & m=n.
\end{cases}
\e
For vacancies, the matrix $\hat K = \hat A + \hat A^T$ has a dimension $N\times
N$ with
\be
\label{vacancy}
A_{nm} = \frac{\exp\big[ \frac{\phi}{2L}(y_n-y_m) -
\frac{i\pi}{4}(\zeta_n-\zeta_m)\big]}
{\sin \big[\frac{\pi}{2L} (\zeta_n x_n+\zeta_m x_m +iy_n - iy_m)\big] },
\e
where $\zeta_n= \pm 1$ if the $n$-th vacancy
belongs to the sublattice A (B).
The analytical expressions (\ref{resonant})--(\ref{vacancy}) can now be used
for the efficient numerical evaluation of the conductance, noise, and higher
transmission cumulants of a disordered graphene sample. Below we focus on
the conductance, Eq.\ (\ref{conductance}). In terms of $\hat K$ it is given by
\be
\label{GfromK}
G = \frac{4e^2}{\pi h} \lt\{
\frac{W}{L} + 2\pi \tr \lt[
(\dot K K^{-1})^2 - \ddot K K^{-1}
\rt]_{\phi=0}
\rt\},
\e
where dots denote derivatives with respect to $\phi$. In the case of resonant
potential impurities, the matrix $K$ is linear in $\phi$; hence the last term
drops from Eq.~(\ref{GfromK}).

Computational efficiency of Eq.~(\ref{GfromK}) is limited by inverting
the matrix 
$K$ at $\phi = 0$. This operation involves $O(N^3)$ multiplications. We run the
standard matrix-inverse update algorithm by adding impurities one by one to
compute the dependence of conductivity on $N$. This reduces the complexity to
$O(N^2)$ per realization on average. The procedure is repeated many times to get
sufficient statistics for different ratios $W/L$. Then we extrapolate the
result to the limit $W \to \infty$, thus eliminating non-universal boundary
effects.

\begin{figure}
\centerline{\includegraphics[width=0.9\columnwidth]{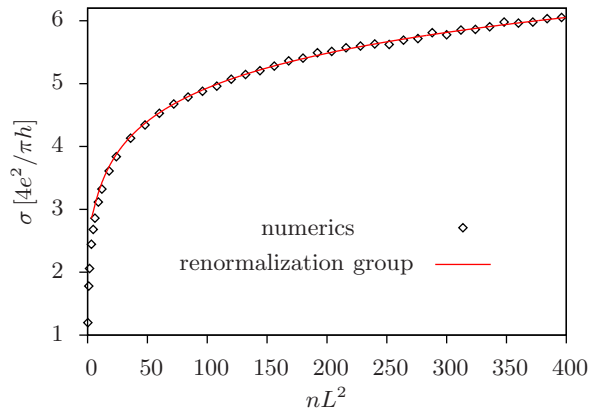}}
\caption{(Color online) Mean conductivity calculated numerically from 
Eqs.\ (\ref{scalar}), (\ref{GfromK}) as a function of concentration
of resonance scalar impurities (data points). The data is obtained by 
interpolation to the limit $W\gg L$. The solid line shows the
result (\ref{sigma-d3}) of the sigma model in class DIII.
}
 \label{fig:scalar}
\end{figure}

For resonant potential impurities, the dependence of the average conductivity
$\sigma = G L/W$ on the impurity concentration $n=N/LW$ in the limit $W\gg L$ 
is plotted in Fig.~\ref{fig:scalar}. To understand this behavior
analytically, we perform the symmetry analysis of the problem. 
The matrix $K$ defined by Eq.~(\ref{scalar})  
yields the Bogoliubov-de-Gennes type of symmetry, $\sigma_x K^T(-\phi)
\sigma_x = - K(\phi)$, which corresponds to the symmetry class D
\cite{Evers08}.  The symmetry of the matrix $K$ is equivalent to that 
of the transfer matrix of the system and can be used to infer the symmetry 
class of the corresponding Hamiltonian, which is given by DIII in the
present case. 
The renormalization group analysis of the corresponding sigma model in
the two-loop approximation yields the following equation 
for the dimensionless conductivity \cite{WZ}
\be
\label{RG}
\frac{d \bar{\sigma}}{d \ln L} = \frac{2}{\pi}\lt(1-\frac{1}{\pi
  \bar{\sigma}}+\mathcal{O}(\bar{\sigma}^{-2})\rt); \ \ \
\bar{\sigma}\equiv \frac{h}{4e^2} \sigma\,,
\e
which holds for $\bar{\sigma}\gg 1$, i.e., in the diffusive
regime. Solving Eq.~(\ref{RG}), we get
\be
\label{sigma-d3}
\sigma =(4e^2/\pi h) \lt(\ln nL^2 - \ln\ln nL^2 \rt).
\e
Figure \ref{fig:scalar} shows a perfect agreement between the
numerical and analytical results.

\begin{figure}
\centerline{\includegraphics[width=0.9\columnwidth]{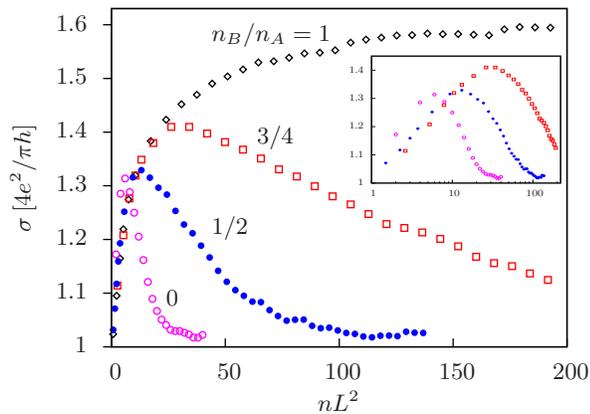}}
\caption{(Color online) Mean conductivity vs. vacancy
  concentration, $n=n_A+n_B$, found 
numerically from Eqs.\ (\ref{GfromK}),
(\ref{vacancy}) for $n_B/n_A = 0, 0.5, 0.75, 1$. Inset illustrates the
conductivity scaling for $n_A \neq n_B$ on the logarithmic scale.
} 
 \label{fig:vacancy}
\end{figure}

An altogether different behavior of the conductivity is obtained  
from Eqs.\ (\ref{vacancy}), (\ref{GfromK})
for graphene with vacancies. The result is plotted in  
Fig.~\ref{fig:vacancy} in the limit $W\gg L$ for different relative
concentrations,
$n_B/n_A =0,1/2,3/4,1$, where $n_A$ and $n_B$ stand
for the vacancy concentrations in the sublattice A and B,
respectively. We see that the conductivity acquires a constant value
for $nL^2\to \infty$. To understand this behavior, we note
that the matrix $K$ from Eq.~(\ref{vacancy}) now possesses  the only
symmetry $K^T=K$ and thus belongs to the symmetry class AI.  
Therefore, graphene with randomly distributed vacancies falls 
into the Hamiltonian symmetry class BDI. The corresponding sigma model is
characterized by a vanishing $\beta$-function, implying 
a constant (in general, nonuniversal) value of conductivity in the
infrared limit. This is
fully consistent with the numerical results of Fig.~\ref{fig:vacancy}.
Remarkably, for  $n_A \ne n_B$, the
 conductivity is a non-monotonic function of $nL^2$ and the limiting
 value is very close to the  $4e^2/\pi h$. 
For equal concentrations, 
 $n_A=n_B$, the numerically obtained limiting value is $\sigma^* \simeq
 1.6 \times 4e^2/\pi h$.   The system is expected to show various
 aspects of criticality characteristic for 2D problem of chiral
 classes \cite{Evers08}.

\textit{Additional comments:} (i) If a small concentration of vacancies is added
to the sample with resonant potential impurities, a crossover from DIII
behavior, Eq.\
(\ref{sigma-d3}), to BDI (saturation) occurs at some high value of
$\bar\sigma$. (ii) We assumed that the inter-impurity distance
is much larger than the graphene lattice constant $a$. For very large
concentration of potential impurities they will start to overlap and
will lose the resonant character. As to vacancies, when their
concentration will increase towards $\sim 1/a^2$, the conductivity
will start to drop, and eventually the system will undergo a
localization transition.

In conclusion, we have developed a theoretical approach to transport
in disordered systems which describes an entire crossover from 
ballistic to diffusive 
or critical regime. The theory can be applied to study localization
physics and criticality in a variety of different systems. 
We have used the theory to calculate the  
conductivity (and, more generally, the full counting statistics)
in undoped graphene with resonant impurities. The 
conductivity increases logarithmically in the case of smooth resonant
potential scatterers (symmetry class DIII) and saturates at a constant value for
vacancies (class BDI). 
In the latter case, the behavior of conductivity depends   
on the vacancy distribution among two sublattices.

We are grateful to F.\ Evers, E.\ Prada, P.\ San-Jose, and A.\ Shytov for
stimulating discussions. The work
was supported by Rosnauka grant 02.740.11.5072 and by
the EUROHORCS/ESF EURYI Award (I.V.G.).


\begin{thebibliography}{99}
\bibitem{WZ} In fact, the sigma model contains an additional
  Wess-Zumino term (in analogy with $\mathbb{Z}_2$ topological term in
  class AII for non-resonant potential scatterers
  \cite{Ostrovsky07a}). However, it does not affect the two-loop
  $\beta$-function at large $\bar{\sigma}$ 
(\ref{RG}) and thus the asymptotic behavior of conductivity
  (\ref{sigma-d3}). The $\beta$-function has a zero at
  $\bar{\sigma}=1/\pi$ corresponding to the Wess-Zumino-Witten fixed
  point. A remarkable property of class DIII is that this fixed point
  is unstable: $\bar{\sigma}$ larger than $1/\pi$ flows to
  infinity.
\end{thebibliography}
\end{document}